# Non-collinear magnetism engendered by a hidden another order

Sergii Khmelevskyi[1*] and Leonid V. Pourovskii[2,3]

[1]*Vienna Scientific Cluster Research Center, Vienna Technical University, Wiedner Hauptstrasse 8-10, 1040, Vienna, Austria*

[2]*CPHT, CNRS, École polytechnique, Institut Polytechnique de Paris, 91120 Palaiseau, France*

[3]*Collège de France, Université PSL, 11 place Marcelin Berthelot, 75005 Paris, France*

**Standard microscopic approach to magnetic orders is based on assuming a Heisenberg form for inter-atomic exchange interactions. These interactions are considered as a driving force for the ordering transition with magnetic moments serving as the primary order parameter. Any higher-rank multipoles appearing simultaneously with such magnetic order are typically treated as auxiliary order parameters rather than a principal cause of the transition. In this study, we show that these traditional assumptions are violated in the case of $PrO_2$. Evaluating the full set of Pr-Pr superexchange interactions from a first-principles many-body technique we find that its unusual non-collinear 2*k* magnetic structure stems from high-rank multipolar interactions, and that the corresponding contribution of the Heisenberg interactions is negligible. The observed magnetic order in $PrO_2$ is thus auxiliary to high-rank "hidden" multipoles. Within this picture we consistently account for previously unexplained experimental observations like the magnitude of exchange splitting and the evolution of magnetic structure in external field. Our findings challenge the standard paradigm of observable magnetic moments being the driving force for magnetic transitions.**

---

[*]Corresponding author, e-mail: sk@cms.tuwien.ac.at

In systems with localized f-electrons, such as compounds with rare-earths or actinides elements, magnetic interactions can induce a rich variety of magnetic orders, often quite complex and non-collinear[1,2]. The origin of these magnetic structures is usually identified as stemming from a competition between on-site anisotropy (crystal field effects) and inter-site interactions coupling the magnetic moments of *f*-ions. However, in general, inter-site exchange interactions also include the terms with higher powers of the on-site magnetic moment operator, namely, multipoles[3]. Discovery of a number of systems with so-called "hidden" order[4,5,6,7] has stimulated growing interest to multipolar super-exchange interactions (SEIs). The "hidden"-order systems undergo a pronounced second-order phase transition without developing any conventional magnetic order directly observable, e.g., in neutron diffraction experiments. Apart from f-electron systems[3,8], heavy transition-metal oxides with strong spin-orbit coupling have been also found to host multipolar orders[9,10,11,12]. However, when an order of sizable conventional magnetic moments is detected experimentally, it is by default assumed to be caused by those moments themselves. Here we show that even in this case multipolar super-exchange interactions can be not only as important as ordinary Heisenberg terms, but that they can entirely determine the magnetic order in a real correlated f-electron system. Our work demonstrates that the experimentally observed magnetic structure of $PrO_2$ is in fact an auxiliary order imposed by an underlying order of high-rank multipolar moments.

The multipolar moment operators, $\hat{O}_L^M$ ($L = 1,..2J, M = -L,..L$) are normalized Stevens operators; the latter have been used in the theory of rare-earth magnetism for over 50 years[1,3,13]. They provide an irreducible operator basis in the space of wave functions for a given ground state multiplet (GSM) of an *f*-ion with total angular momentum $J$ (see SI Table S1 for the list of Stevens operators up to $L = 5$). Any Hamiltonian acting within the GSM of an f-ion can be expressed as a linear combination of the $\hat{O}_L^M$ operators. Therefore, the Hamiltonian describing a system of interacting ions with localized f-electron shells generally reads:

$$\hat{H} = \sum_i \hat{H}_{CF}^i + \frac{1}{2} \sum_{i,j} \sum_{LML'M'} V_{LM,L'M'}^{ij} \hat{O}_{L\ (i)}^M \hat{O}_{L'(j)}^{M'}, \qquad (1)$$

where the summations run over the $i,j$ lattice sites. The parameters $V_{LM,L'M'}^{ij}$ are two-site super-exchange interaction (SEI) parameters and $\hat{H}_{CF}^i = \sum_{LM} B_{LM} \hat{O}_L^M$ is the standard single-site CF Hamiltonian[1] expressed through unnormalized Stevens operators $\hat{O}_L^M$. The number of higher rank multipoles (*L > 1*) and their associated interaction parameters increase rapidly with the multiplicity of the ground state of the f-ion.

Unless there is experimental evidence of "hidden" multipolar order in the system or if large lattice distortions are associated with the magnetic ordering, only first-order (dipole) inter-site interaction terms with *L*=1 are typically considered in theoretical analysis. They are of the handbook[1] Heisenberg-like form $\sum_{i,j} \hat{J}_i \bar{I}_{ij} \hat{J}_j$, where the dipole magnetic moment at the site *i* is given by $g\mu_B \hat{J}_i$, $\bar{I}_{ij}$ is 3x3 interaction matrix, $g_J$ is the Lande factor and $\mu_B$ is the Bohr magneton. The usual theoretical approach is then to look for the interaction matrices $\bar{I}_{ij}$ that stabilize of the observed magnetically ordered structure[14], considered as the prime order. The non-zero expectation values of the higher order multipoles that result from the prime order are auxiliary order parameters. In this paradigm, they are expected to provide small corrections to the Heisenberg Hamiltonian or to the Landau expansion of the free energy.

In this study, we identify a system - the $PrO_2$ dioxide - in which high-rank multipolar interactions are responsible for occurrence of the magnetic transition and stabilization of the observed magnetic order. We use first-principle based many-body techniques to calculate the full set of multipolar SEI and then solve the ab initio effective many-body Hamiltonian (1).

At high temperatures, the $PrO_2$ compound has a fluorite crystal structure similar to the dioxides of actinides[15]. At T = 120 K, it undergoes a Jahn-Teller (JT) structural transition into a structure where the oxygen atoms are shifted from their initial positions in the fluoride structure[16,17,18], as shown in Fig. 1a. The resulting oxygen "chiral" structural order doubles the fluorite cubic unit cell in the direction perpendicular to the plane of the oxygen shifts[16]. In the high temperature phase, the ground state multiplet (GSM) $^2F_{5/2}$ (J=5/2) of the $Pr^{4+}$ Kramer's ions is split by the cubic crystal field into ground state quartet and doublet states (~130 meV). Below 120 K the JT effect further splits the ground state quartet into two doublets, with a broad excited level centered at ~30 meV at low temperatures[19]. At the Néel temperature $T_N$ = 13.5 K, a complex non-collinear antiferromagnetic (AFM) order sets in[17], as shown in Fig. 1b. The magnetic unit cell coincides with the distorted crystal unit cell. The orientation of the magnetic moments in the ordered state is parallel to the oxygen shifts, and there are four non-equivalent magnetic planes. The structure can be better understood by considering separately the Pr magnetic moment components parallel to the edges of the crystal lattice[17], as shown in Fig. 3d. At each Pr site, the moduli of these components are the same. The AFM order of the large component (red arrows in Fig. 3d) is just the alternating AFM order of ferromagnetically ordered basal planes that corresponds to the propagation vector $k$=[001]. The small component (blue) exhibits a simple AFM order in the basal plane that repeats itself every fourth layer ($k$=[10$\frac{1}{2}$]). It has been shown that the JT effect and the observed CF splitting can be readily understood by employing a standard CF model (1) [19,20]. Moreover, it has also been demonstrated that direct total energy calculations within the framework of ab-initio DFT+U methodology can predict the stabilization of the observed distorted structure[21]. However, the physical mechanism behind the observed magnetic order (Fig. 1b) remains unclear.

Jensen[20] has shown that this magnetic order can be stabilized by including three competing Heisenberg interactions of the same order of magnitude between the Pr ions on three distant nearest neighbor (NN) atomic shells. However, the Heisenberg-like model of Ref. 20 is not able to consistently account for the magnitude of exchange splitting of the lowest CF doublet in the ordered phase[19]. Neither the changes of magnetic structure and structural domain populations under applied field are consistently explained within its framework. Moreover, this model includes a large and ferromagnetic 2nd NN SEI. For a magnetic insulator, where the SEIs are expected to be short-ranged and predominantly AFM in character, such a picture seems quite unlikely. As shown below, our calculations do predict the 2nd and 3rd NN SEIs to be negligible compared to the NN ones.

To calculate the full set of parameters required for the effective many-body Hamiltonian (1) for $PrO_2$, we first obtained the paramagnetic CF splitting and composition of the corresponding wave functions for both the cubic fluorite and experimental JT distorted structures using a charge self-consistent DFT + Dynamical Mean-Field Theory (DMFT) technique[22,23,24]. We treat many-body effects within the Pr 4f shell in the quasi-atomic Hubbard-I approximation (see Methods). Fig. 2a shows the ab-initio-calculated splitting of the energy levels; we obtain 33 and 103 meV for the first and second excited doublets, respectively, in the JT distorted structure. It is in good agreement with the experimental values of ~30 and 130 meV, respectively[19]. The obtained CF splitting and corresponding wave functions, which are listed the Supplemental Material (SM), define the single-site CF part of the Hamiltonian (1).

The interaction term in equation (1) includes 35 multipolar operators for the Pr$^{4+}$ ion in the GSM (J=5/2) configuration (listed in SI Table S1). To obtain the full 35x35 interaction matrix $V_{LM,L'M'}^{ij}$, we used the force theorem in the Hubbard-I Approximation (FT-HI)[25], see Method for details. Only the NN SEIs are important, the second NN SEIs are at least an order of magnitude smaller, the longer range SEIs are negligible. The calculated NN SEI is presented in graphic form in Fig. 2b, with corresponding numerical values given in the SI. Since the interaction between multipolar moments of different parity (with odd and even L number) must be zero due to time-reversal symmetry, the interaction matrix has a block-wise form. However, the number of symmetry-independent non-zero elements is still quite large, rendering a phenomenological estimation of the SEI quite unfeasible. This is one of the main reasons why contemporary theories of magnetism are typically limited to the Heisenberg interaction term.

Using ab-initio-derived parameters of the full interacting Hamiltonian (eq. 1), we apply a generalized mean-field approximation (MFA) to study magnetic and multipolar phase transitions in PrO$_2$. To cover all possible orderings within the crystal cell of the distorted structure (Fig. 1), we introduce eight different Pr sublattices (see Method for details). The self-consistent solution of the MFA equations yields a second order phase transition at T$_N$ = 21 K where new non-zero order parameters $\langle \hat{O}_L^M{}_{(i)} \rangle$ emerge. The temperature dependence and schematic pictures of these order parameters are shown in Fig. 3.

In the basal planes of the crystal structure there are two components of the magnetic moment: $m_x = \langle \hat{O}_1^1 \rangle$, $m_y = \langle \hat{O}_1^{-1} \rangle$ (Fig. 3a). There are also four octupolar moment components $\langle \hat{O}_3^{-3,-1,1,3} \rangle$, and five triakontadipolar components $\langle \hat{O}_5^{-3,-1,1,3,5} \rangle$ (see Fig. 3b and 3c, respectively), resulting in a combined AFM order of odd rank multipoles. The mutual orientation of the magnetic moments in different sub-lattices in the self-consistent solution exactly matches the experimental magnetic order presented in Fig. 1b. Thus, the calculated 1NN multipolar super-exchange interactions stabilize the experimental magnetic structure at low temperatures. The derived value of the Neel temperature is overestimated compared to the experimental value[15] of 13.5 K, mainly due to the MFA approximation. It is known that MFA produces much higher T$_N$ values in cases of competing AFM interactions on the geometrically frustrated fcc lattice. The zero temperature MFA magnetic moments $m_x = 0.49 \mu_B$, $m_y = 0.82 \mu_B$ are in fair agreement to the neutron diffractions estimates[17] of $m_x = 0.35 \mu_B$, $m_y = 0.68 \mu_B$. Our MFA magnetic moments are better agreement with experiment as compared to those calculated by Jensen[20] from a semi-empirical Heisenberg Hamiltonian. The MFA moments are expected to be somewhat overestimated since we neglect the dynamical Jahn-Teller effect, which tend to reduce the atomic moments in PrO$_2$ compound[19]; this reduction was estimated by Ref. 20 to be about 5%. A similar reduction of the magnetic moments due to the dynamical Jahn-Teller effect is also characteristic for the isostructural UO$_2$ compound[26]. The magnetic order in PrO$_2$ induces an exchange splitting of the CF doublet with the magnitude of 2.7 meV at $T$=10 K (75% of experimental $T_N$) as measured by inelastic neutron scattering[19]. The Heisenberg model of Jensen[20] underestimated this splitting by one-third, which was unexpected provided that the same model overestimated the experimental magnetic moments by about 40%. We calculated the splitting of Pr $J$=5/2 levels in the magnetically ordered phase using the converged mean field. We obtained the splitting of the ground state doublet (2.64 meV) that matches the experiment at the correspondingly scaled temperature of 15 K (about 75% of theoretical $T_N$). This splitting is predicted by our calculations to reach 3.43 meV at zero temperature.

PrO$_2$ also exhibits a complex evolution of the magnetic structure and structural domain distribution under applied magnetic field. An initial application of 5 Tesla field $H$ along the [110] cubic direction was found[18] to irreversibly change the distribution of structural domains corresponding to symmetry equivalent JT distortion patterns. Subsequent application of the field in the magnetically ordered phase led to drastic changes in the intensities of magnetic Bragg reflections, which signal significant modifications in the magnetic structure of the preferable structural domain. This field evolution remains poorly understood, since it could not be accounted for within the Heisenberg picture[20]. We studied the evolution of magnetic structure under $H||<110>$ by supplementing the calculated Hamiltonian with the corresponding Zeeman term and solving it in mean-field. We find the field lying in the basal $ab$ plane of the JT structure (Fig. 1a) to be energetically preferable. The calculated magnetic structure under the field $H$=5 Tesla is highly unusual (Fig. 1c). The moments of the Pr sites that in the initial structure formed large angles with the field direction keep their large-component propagation vector $k$=<001> and are aligned orthogonally to the field. The Pr moments initially forming small angles with the field direction are reduced by half and aligned alone the $a$ or $b$ cube edges. This magnetic structure is much more complex than the one initially suggested[18] to account for the experimental picture. We find, however, that the predicted structure (Fig. 1c) accounts well for the measured field evolution of Bragg intensities[18]. Using the experimental setup (i. e. assuming $H||[0\bar{1}1]$ as in Ref. 18) we calculated the intensities of the Bragg peaks measured in Ref. 18. In agreement with those measurements, the (011) Bragg peak is predicted to disappear under applied field, while the (100) peak intensity is almost unchanged and the (211) one diminishes by about 10% (see SM Table S3). We note, however, that our calculations predict a too rapid decay of the (011) reflex with increasing field compared to experiment; this might stem from additional anisotropy contribution due to the dynamic JT effect being neglected in our calculations.

As shown above, the full calculated SEI Hamiltonian (1) provides a faithful description for the properties of magnetically ordered PrO$_2$. Since a very large number terms are included into this Hamiltonian, it is important to understand the impact of different interactions on the complex magnetic and multipolar order in PrO$_2$. To that end, we solved MFA equations only with standard Heisenberg magnetic interactions (dipolar interactions) included. Thus, we used only the dipole-dipole block (L=1) of calculated SEI, $\sum_{i,j}\sum_{MM'} V^{ij}_{1M,1M'} \hat{O}^M_{1(i)} \hat{O}^{M'}_{1(j)}$, disregarding all higher order multipolar interactions. This can be recast in the standard form $\sum_{i,j} \hat{J}_i \bar{I}_{ij} \hat{J}_j$ of a Heisenberg-like interaction between dipole magnetic moments. As expected from earlier analyses provided by Jensen[20], solving this Heisenberg-like model in MFA, we derived a wrong AFM order. More importantly, the calculated Neel temperature was found to be just 0.2 K, two orders of magnitude smaller than the ordering temperature obtained with the full calculated SEI matrix.

Thus, the dominant contribution to the ordering in $PrO_2$ must be due to high-order multipolar SEIs. We verified this explicitly, by setting in eq. 1 all SEIs involving dipole operators to zero. By solving eq. 1 with the dipole-moment contribution completely suppressed, we find essentially the same magnetic structure with the magnitude of magnetic moment reduced by about 6% and $T_N$=18 K. Hence, the actual contribution of dipole moments into the formation of ordered structure is insignificant. This can be further confirmed by evaluating the contributions of various SEI, $V^{ij}_{LM,L'M'} \langle \hat{O}^M_L \rangle_i \langle \hat{O}^{M'}_{L'} \rangle_j$, into the magnetic ordering energy; this analysis shows that the leading contributions are due to octupolar and trikontadipolar SEI (see SI Table S2). This means that the observed magnetic order in $PrO_2$ is in fact induced by multipolar interactions, with the actual contribution by dipole moments being quite negligible.

To understand the mechanism through which the auxiliary dipole order is induced we determined among the J=5/2 moments those that are active within the lowest CF doublet. Since the JT splitting of 30 meV between this doublet and the first excited one (Fig. 2a) is an order of magnitude larger than the exchange splitting of about 3 meV, one may assume that the admixture of excited CF levels by the exchange field is negligible. The degrees-of-freedom of the lowest CF doublet can be represented by a pseudo-spin-1/2 variable $\tau$ with the two Kramers' partners being the two eigenstates of $\tau_z$. The physical content of these four spin-1/2 operators $\tau_\alpha$ – the identity $\tau_I$ as well as $\tau_x, \tau_y$, and $\tau_z$ – is determined by upfolding them into the full J=5/2 as $\tilde{\tau}_\alpha = P \tau_\alpha P^\dagger$, where the columns of the projection matrix $P$ are the CF doublet eigenstates in the J=5/2 basis. Subsequently expanding $\tilde{\tau}_\alpha$ into the J=5/2 Stevens operators $\hat{O}^M_L$ one finds

$$\tilde{\tau}_x = 0.119 \hat{O}^x_1 - 0.237 \hat{O}^y_1 + 0.467 S^x_3 + 0.460 S^x_5,$$

$$\tilde{\tau}_y = -0.247 \hat{O}^x_1 + 0.139 \hat{O}^y_1 + 0.471 S^y_3 + 0.445 S^y_5,$$

where for brevity we introduce a shorthand notation $S^\alpha_3$ ($S^\alpha_5$) for normalized superpositions of octupole (trikontadipole) Stevens operators contributing to a given $\tau_\alpha$. Analogous expression is found for $\tau_z$, which includes only z-projections of the magnetic operators and thus, due to a strong planar anisotropy, plays no role in the magnetic order. The identity $\tau_I$ is formed by time-even J=5/2 operators corresponding to the CF; thus, it also plays no role.

One sees that any doublet-space density matrix for an arbitrary planar order, $\rho = \tau_x \langle \tau_x \rangle + \tau_y \langle \tau_y \rangle$, will include about the same total contributions due to dipoles, octupoles and trikontadipoles, since their respective contributions to both $\tau_x$ and $\tau_y$ are very similar. The strong JT splitting thus entangles magnetic operators of different ranks together; this is how dominating multipolar SEIs induce the auxiliary dipole order in $PrO_2$. Due to this entanglement, separating the order parameters presented in Fig. 3 into primary and secondary ones has no physical significance. However, one still may ask the physically relevant question: which interactions drive the magnetic order of PrO$_2$? Our direct calculations convincingly show that those interactions are multipolar. Among various multipolar order parameters shown in Fig. 3, the octupoles $\hat{O}_3^1$ ($xz^2$) and trikontadipoles $\hat{O}_5^1$ ($xz^4$), which correspond to a magnetic density oblate along the secondary component ($y$) direction, provide the leading contribution to the ordering energy (see SI Table S2).

In general, there is no physical reason why dipolar inter-site exchange interactions should dominate higher-order multipolar ones in heavy ions with unquenched orbital moments. This work demonstrates that the prevailing paradigm, which suggests that observed magnetic order arises solely from interactions between the observed order parameters, can be violated in real systems. A possible prominent role of the higher order multipolar interactions in PrO$_2$ has been suggested previously[3,20]. However, here, we show that those multipolar SEI are not simply significant, but that they fully determine the physics of the magnetic behavior in PrO$_2$. Complex combined order parameters entangling both magnetic and multipolar interactions, uncovered here in the case of PrO$_2$, could be rather common in localized f-electron systems. Consequently, the analysis of magnetic structures in f-electron materials should incorporate multipolar degrees of freedom from the onset, even in cases where only dipolar moments are experimentally detected.

**Method**

*Ab initio calculations.* We calculated the electronic structure of PrO$_2$ within the DFT+DMFT framework using the experimental lattice structures of both the high-temperature cubic structure and the "chiral" JT-distorted body-centered tetragonal one[17]. In these calculations we employed the full-potential DFT code Wien2k[28] and the charge self-consistent DMFT implementation provided by "TRIQS" library[24,29]. We treat Pr 4f states within the Hubbard-I (HI) approximation. In these calculations, we employ the Wien-2k basis cutoff $R_{MT}K_{max}$=8, local density approximation, and include spin-orbit. 3000 and 800 ***k***-points in the full Brillouin zone were employed in the cubic and tetragonal cases, respectively. The on-site Coulomb repulsion for Pr 4*f* shell was parametrized by *U*=*F*$^0$=6.5 eV and *J*$_H$=0.73 eV; these values agree with previous applications of DFT+HI calculations to Pr compounds[30,31]. The projective 4f Wannier orbitals were constructed from the 4f-like Kohn-Sham (KS) bands within the range [-1:2.7] eV with respect to the KS Fermi level. The fully localized limit double counting was calculated using the nominal *N*=1 occupancy of Pr 4f.

The CF splitting and states for the Pr 4f shell were obtained from converged DFT+HI calculations. The SEIs for all Pr-Pr bonds within several first correlation shells were obtained using the FT-HI method[25] applied as a postprocessing of the converged DFT+HI electronic structure of the cubic structure. We previously used the FT-HI method to calculate super-exchange interactions for the actinide dioxides UO$_2$ and NpO$_2$ [27,28]. Only the SEIs coupling the lowest CF levels in the cubic structure - triplet and quartet, respectively – were calculated for those systems. We found this strategy to be not appropriate here since the JT distortion is rather large in PrO$_2$ strongly mixing the cubic GS $\Gamma_8$ and excited $\Gamma_6$ levels. Therefore, we evaluated the SEI interaction matrix $V^{ij}_{LM,L'M'}$ for the full J=5/2 multiplet and explicitly kept the CF term in the Hamiltonian (1).

*Mean-field calculations.* We solved the many-body effective Hamiltonian (1) in the MFA considering all possible orders realizable within the distorted unit cell of PrO2 (Fig. 1a). Introducing eight different Pr sublattices for a mean-field Hamiltonian

$$\hat{H}_{mf} = \sum_i \hat{H}^i_{CF} + \sum_{i,j(NN)} \sum_{LM} V^{ij}_{LM,L'M'} \hat{O}^M_{L\ (i)} \langle \hat{O}^{M'}_{L'(j)} \rangle, \quad (2)$$

we obtained a set of coupled mean-field equations for each of the 35 multipolar moments $\langle \hat{O}^M_{L\ (i)} \rangle$ on each $i^{th}$ sub-lattice, resulting in a system of 280 equations:

$$\langle \hat{O}^M_{L\ (i)} \rangle = \frac{Tr\left[\hat{O}^M_{L\ (i)} e^{-\hat{H}_{mf}/kT}\right]}{Tr\left[e^{-\hat{H}_{mf}/kT}\right]} \quad \begin{array}{c}(i=1,\ldots 8)\\(L=1,\ldots,2J, M=-L,\ldots,L)\end{array} \quad (3)$$

which we solve iteratively.

*Calculations of neutron-scattering intensities.* We calculated the intensities of neutron Bragg peaks using the beyond-dipole-approximation approach of ref. 11. Namely, the intensity of elastic scattering $I(\boldsymbol{q})$ at the scattering vector $\boldsymbol{q}$ reads

$$I(\boldsymbol{q}) = \sum_{\alpha\beta}(\delta_{\alpha\beta} - \hat{q}_\alpha \hat{q}_\beta) \sum_{\mu\nu} F_{\mu\alpha}(\boldsymbol{q}) F_{\nu\beta}(\boldsymbol{q}) \langle O_\mu(\boldsymbol{q}) \rangle \langle O_\nu(\boldsymbol{q}) \rangle,$$

where $\langle O_\mu(\boldsymbol{q}) \rangle = \sum_i \langle O_\mu \rangle_i e^{i\boldsymbol{q}R_i}$ is the Fourier transform of the order parameter $\langle O_\mu \rangle_i$ for the magnetic multipole $\mu \stackrel{\text{def}}{=} [LM]$ at site $i$, $\boldsymbol{R}_i$ is the coordinate of this site within the magnetic unit cell, $F_{\mu\alpha}(\boldsymbol{q})$ is the form-factor for the multipole $\mu$ and $\alpha = x, y, \text{or } z$, $\hat{\boldsymbol{q}}=\boldsymbol{q}/|\boldsymbol{q}|$. The form-factors $F_{\mu\alpha}(\boldsymbol{q})$ were calculated by evaluating matrix elements of the neutron scattering operator[32] in the Pr J=5/2 basis and expanding the resulting matrix in the basis of J=5/2 magnetic multipole operators[33] as detailed in Supplementary of ref. 11. In calculating the matrix elements, we used the radial integrals $\langle j_l \rangle$ calculated[34] for $Pr^{3+}$, since this data for $Pr^{4+}$ are not available in the literature. In spite of the large magnitudes of high-rank multipolar order parameters in $PrO_2$, their contribution to the Bragg peaks intensities that we calculated was found to be negligible due to their small form-factors.

**Competing Interests:**
The authors declare no competing interests.

**Acknowledgments:**
L.V.P. acknowledges support from European Research Council Grant ERC-319286-"QMAC" and the computer team at CPHT.

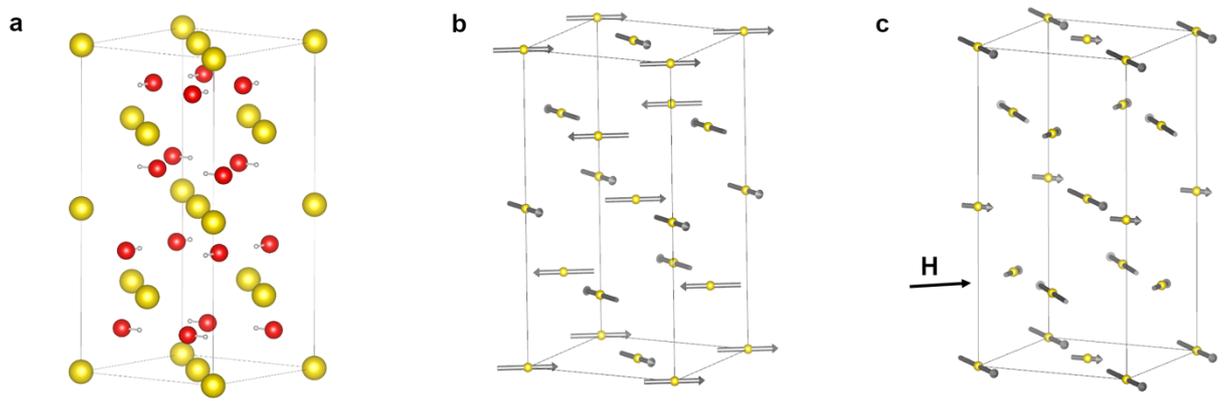

**Fig. 1. Lattice and magnetic structure of PrO$_2$.** a) The distorted crystal structure below 120 K. Yellow balls are Pr, the oxygens positions in the ideal cubic fluorite structure (small white circles) and the corresponding oxygen positions in the distorted structure (red balls) are connected by lines representing the shift due to the distotion (the shift's magnitude is enhanced with respect to the experimental one for clarity); b) Calculated antiferromagnetic order at 1.5 K (only Pr atoms are shown for clarity); c) Predicted antiferromagnetic order in PrO$_2$ under external magnetic field of 5 Tesla applied in the [110] direction (indicated in the plot by the black arrow).

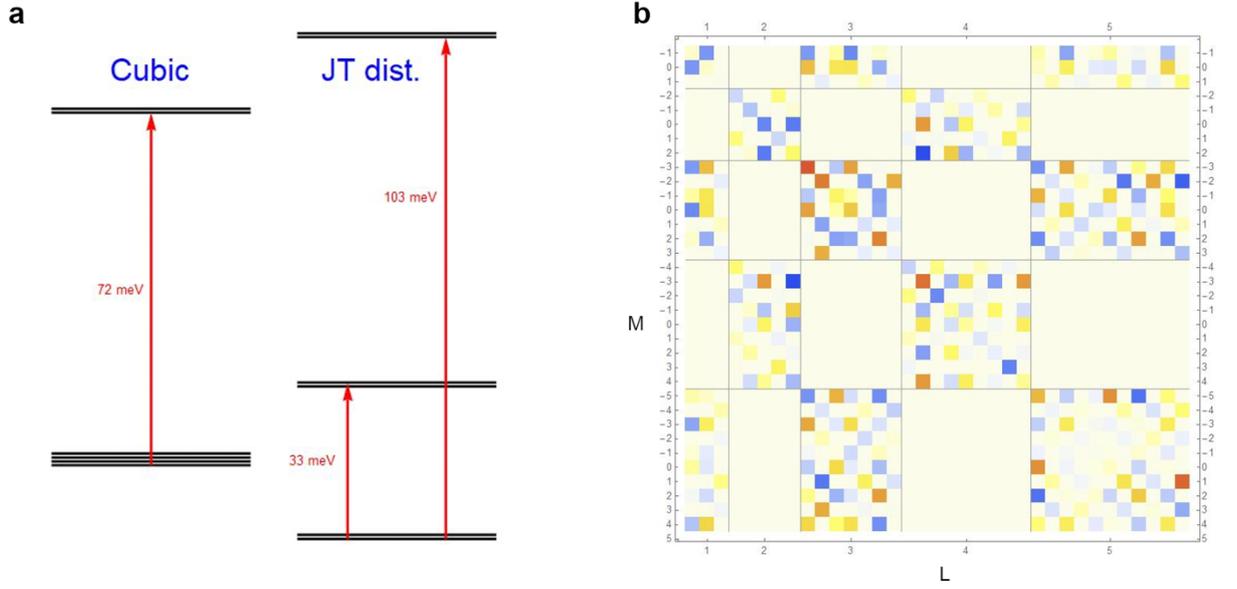

**Fig. 2. Crystal-field splitting and intersite exchange interactions in PrO$_2$.** a). Calculated CF splliting of the ground state $4f^1$ multiplet of the Pr$^{4+}$ ion in the crystal field of PrO$_2$ in the cubic high temperature phase (left) and in the experimental Jahn-Teller distorted phase (right) at low temperatures; b). Temperature map of the calculated super-exchange interaction matrix $V^{ij}_{LM,L'M'}$, see eq. 1, for the next-neighbor Pr-Pr fluorite structure bond ($R$=[0, 1/2, -1/2]) in the corresponding Cartesian system. Top/Botom ticks are L indices (multipoles ranks). Left/Right ticks indicate the projections M within each corresponding L-blocks. Warm colors mark ferromagnetic (negative) and cold colors antiferromagnetic (positive) superexchange interactions. The corresponding numerical values are given in the Suplementary Information.

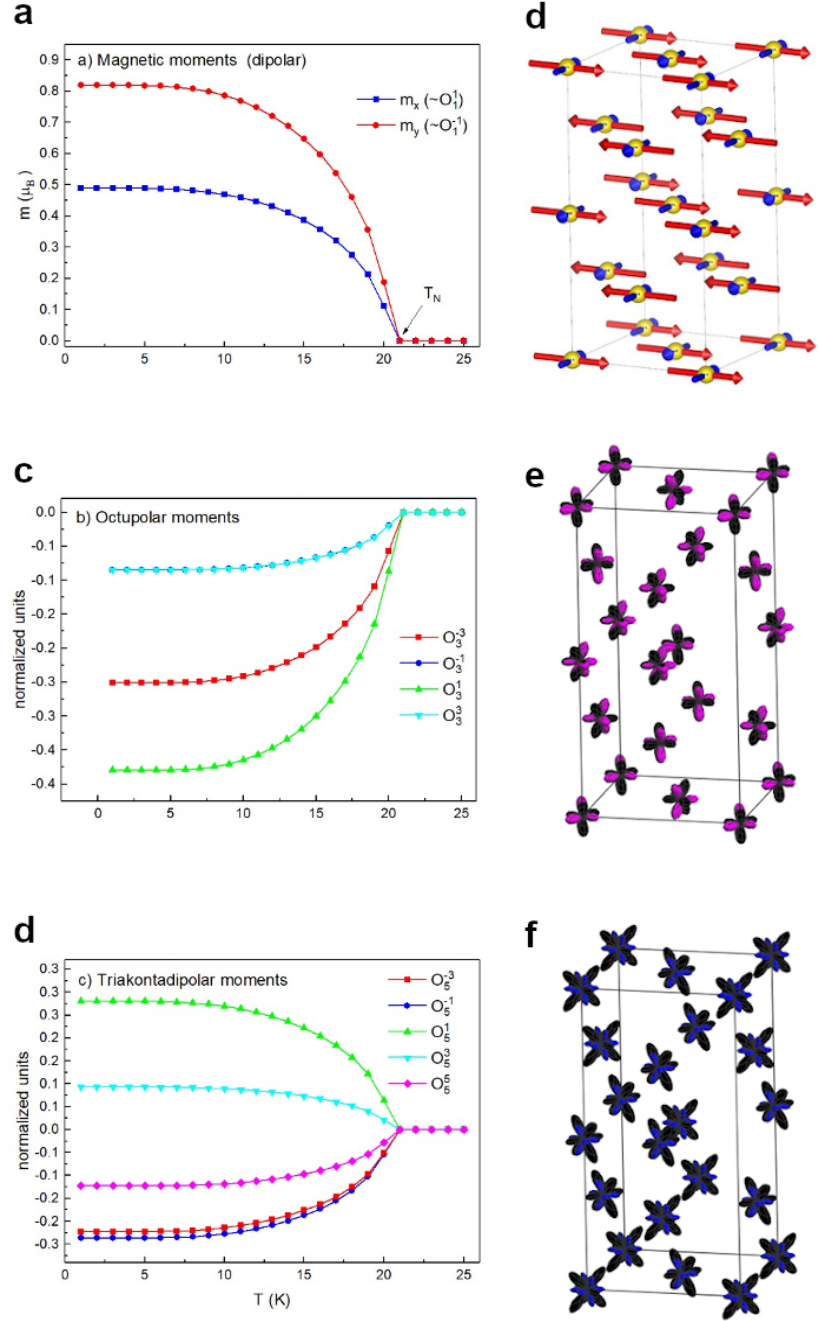

**Fig. 3. Magnetic multipolar order in PrO$_2$.** Temperature dependences of the order parameters obtained by solving eq. 1 in mean-field are shown in (a-c); the corresponding ordered moments are schematically depicted in (d-f). (a). The magnitudes of the two dipole-moment components – principal $M_x$ (propagation vector $\mathbf{k}$=[001]) and secondary $M_y$ ($\mathbf{k}$=[10$\frac{1}{2}$]) – are indicated by red and blue colors, respectively. The corresponding magnetic structure is shown in panel (d). The magnitude of magnetic octupolar and triakontadipolar on-site moments vs. temperature is shown in panels (b) and (c); the moments are separated into non-zero projections. The corresponding total octupolar and triakontadipolar moments are schematically depicted in panels (e) and (f), respectively. The moment expectation values are calculated using the normalized form of the moment operators; the size of moments depicted in panels (e) and (f) is proportional to their absolute magnitude.

# References.

# Supplementary Information

## for " Non-collinear magnetism engendered by a hidden another order"

by Sergii Khmelevskyi and Leonid V. Pourovskii

### 1. Stevens operators

In Table S1 we list the operator forms for the unnormalized Stevens operators $\hat{\mathcal{O}}_L^M$ up to $L=5$. The normalized operators $\hat{O}_L^M$ are obtained from the unnormalized ones as $\hat{O}_L^M = \frac{1}{\sqrt{N}}\hat{\mathcal{O}}_L^M$, where the norm $N = Tr\,[\hat{\mathcal{O}}_L^M \cdot \hat{\mathcal{O}}_L^M]$.

| rank | Notation | Operator form in $|JM_J\rangle$ basis |
|---|---|---|
| **Dipoles Rank 1** | $\hat{\mathcal{O}}_1^{-1}$ | $\hat{J}_y$ |
| | $\hat{\mathcal{O}}_1^0$ | $\hat{J}_z$ |
| | $\hat{\mathcal{O}}_1^1$ | $\hat{J}_x$ |
| **Quadrupoles Rank 2** | $\hat{\mathcal{O}}_2^{-2}$ | $\overline{\hat{J}_x\hat{J}_y} = \hat{J}_x\hat{J}_y + \hat{J}_y\hat{J}_x$ |
| | $\hat{\mathcal{O}}_2^{-1}$ | $\overline{\hat{J}_y\hat{J}_z} = \hat{J}_y\hat{J}_z + \hat{J}_z\hat{J}_y$ |
| | $\hat{\mathcal{O}}_2^0$ | $3\hat{J}_z^2 - J(J+1)$ |
| | $\hat{\mathcal{O}}_2^1$ | $\overline{\hat{J}_x\hat{J}_z} = \hat{J}_x\hat{J}_z + \hat{J}_z\hat{J}_x$ |
| | $\hat{\mathcal{O}}_2^2$ | $\hat{J}_x^2 - \hat{J}_y^2$ |
| **Octupoles Rank 3** | $\hat{\mathcal{O}}_3^{-3}$ | $3\overline{\hat{J}_x^2\hat{J}_y} - \hat{J}_y^3$ |
| | $\hat{\mathcal{O}}_3^{-2}$ | $\overline{\hat{J}_x\hat{J}_y\hat{J}_z}$ |
| | $\hat{\mathcal{O}}_3^{-1}$ | $5\overline{\hat{J}_y\hat{J}_z^2} - J(J+1)\hat{J}_y$ |
| | $\hat{\mathcal{O}}_3^0$ | $5\hat{J}_z^3 - 3J(J+1)\hat{J}_z$ |
| | $\hat{\mathcal{O}}_3^1$ | $5\overline{\hat{J}_x\hat{J}_z^2} - J(J+1)\hat{J}_x$ |
| | $\hat{\mathcal{O}}_3^2$ | $\overline{\hat{J}_x^2\hat{J}_z} - \overline{\hat{J}_y^2\hat{J}_z}$ |
| | $\hat{\mathcal{O}}_3^3$ | $\hat{J}_x^3 - 3\overline{\hat{J}_x\hat{J}_y^2}$ |
| **Hexadecapoles Rank 4** | $\hat{\mathcal{O}}_4^{-4}$ | $\overline{\hat{J}_x^3\hat{J}_y} - \overline{\hat{J}_x\hat{J}_y^3}$ |
| | $\hat{\mathcal{O}}_4^{-3}$ | $3\overline{\hat{J}_x^2\hat{J}_y\hat{J}_z} - \overline{\hat{J}_y^3\hat{J}_z}$ |
| | $\hat{\mathcal{O}}_4^{-2}$ | $7\overline{\hat{J}_x\hat{J}_y\hat{J}_z^2} - J(J+1)\overline{\hat{J}_x\hat{J}_y}$ |
| | $\hat{\mathcal{O}}_4^{-1}$ | $7\overline{\hat{J}_y\hat{J}_z^3} - 3J(J+1)\overline{\hat{J}_y\hat{J}_z}$ |
| | $\hat{\mathcal{O}}_4^0$ | $35\hat{J}_z^4 - 30J(J+1)\hat{J}_z^2 + 3J^2(J+1)^2$ |
| | $\hat{\mathcal{O}}_4^1$ | $7\overline{\hat{J}_x\hat{J}_z^3} - 3J(J+1)\overline{\hat{J}_x\hat{J}_z}$ |
| | $\hat{\mathcal{O}}_4^2$ | $7\overline{\hat{J}_x^2\hat{J}_z^2} - 7\overline{\hat{J}_y^2\hat{J}_z^2} - J(J+1)(\hat{J}_x^2 - \hat{J}_y^2)$ |
| | $\hat{\mathcal{O}}_4^3$ | $\overline{\hat{J}_x^3\hat{J}_z} - 3\overline{\hat{J}_x\hat{J}_y^2\hat{J}_z}$ |
| | $\hat{\mathcal{O}}_4^4$ | $\hat{J}_x^4 - 6\overline{\hat{J}_x^2\hat{J}_y^2} + \hat{J}_y^4$ |
| **Triakontadipoles Rank 5** | $\hat{\mathcal{O}}_5^{-5}$ | $5\overline{\hat{J}_x^4\hat{J}_y} - 10\overline{\hat{J}_x^2\hat{J}_y^3} + \hat{J}_y^5$ |
| | $\hat{\mathcal{O}}_5^{-4}$ | $\overline{\hat{J}_x\hat{J}_y^3\hat{J}_z} - \overline{\hat{J}_x^3\hat{J}_y\hat{J}_z}$ |
| | $\hat{\mathcal{O}}_5^{-3}$ | $-27\overline{\hat{J}_x^2\hat{J}_y\hat{J}_z^2} + 9\overline{\hat{J}_y^3\hat{J}_z^2} + J(J+1)\left(3\overline{\hat{J}_x^2\hat{J}_y} - \hat{J}_y^3\right)$ |
| | $\hat{\mathcal{O}}_5^{-2}$ | $-3\overline{\hat{J}_x\hat{J}_y\hat{J}_z^3} + J(J+1)\overline{\hat{J}_x\hat{J}_y\hat{J}_z}$ |
| | $\hat{\mathcal{O}}_5^{-1}$ | $21\overline{\hat{J}_y\hat{J}_z^4} - 14J(J+1)\overline{\hat{J}_y\hat{J}_z^2} + J^2(J+1)^2\hat{J}_y$ |
| | $\hat{\mathcal{O}}_5^0$ | $63\hat{J}_z^5 - 70J(J+1)\hat{J}_z^3 + 15J^2(J+1)^2\hat{J}_z$ |
| | $\hat{\mathcal{O}}_5^1$ | $21\overline{\hat{J}_x\hat{J}_z^4} - 14J(J+1)\overline{\hat{J}_x\hat{J}_z^2} + J^2(J+1)^2\hat{J}_x$ |

|   |   |
|---|---|
| $\hat{O}_5^2$ | $3\left(\overline{\hat{J}_x^2\hat{J}_z^3} - \overline{\hat{J}_y^2\hat{J}_z^3}\right) - J(J+1)\left(\overline{\hat{J}_x^2\hat{J}_z} - \overline{\hat{J}_y^2\hat{J}_z}\right)$ |
| $\hat{O}_5^3$ | $27\overline{\hat{J}_x\,\hat{J}_y^2\hat{J}_z^2} - 9\overline{\hat{J}_x^3\hat{J}_z^2} + J(J+1)\left(\hat{J}_x^3 - 3\overline{\hat{J}_x\,\hat{J}_y^2}\right)$ |
| $\hat{O}_5^4$ | $\overline{\hat{J}_x^4\hat{J}_z} - 6\overline{\hat{J}_x^2\hat{J}_y^2\hat{J}_z} + \overline{\hat{J}_y^4\hat{J}_z}$ |
| $\hat{O}_5^5$ | $\hat{J}_x^5 - 10\overline{\hat{J}_x^3\hat{J}_y^2} + 5\overline{\hat{J}_x\,\hat{J}_y^4}$ |

Table S1: Unnormalized Stevens operators in terms of the total angular momentum operators.

## 2. Superexchange interactions (SEIs) in PrO$_2$

Below we list the full SEI matrix $V_{LM,L'M'}^{ij}$ for the nearest-neighbor lattice vector $\mathbf{R}=[0,1/2,-1/2]$, which is schematically shown in Fig. 2b of the main text as a color map. Since the interaction matrix is symmetric, we show only one of the two identical off-diagonal blocks coupling $L$ and $L'$ multipoles. We first list the interaction matrices for the magnetic multipoles (odd $L$) and then those for the charge multipoles (even $L$). First row/column indicate $M'/M$, respectively. All values are in meV, the matrix elements below 0.01 meV in absolute value are omitted.

**Magnetic SEIs**

**Dipole-dipole ($L=1$, $L'=1$)**

|    | -1   | 0    | +1   |
|----|------|------|------|
| -1 | 0.04 | 0.35 |      |
| 0  | 0.35 | 0.04 |      |
| 1  |      |      | -0.01 |

**Octupole-octupole ($L=3$, $L'=3$)**

|    | -3    | -2    | -1    | 0     | +1    | +2    | +3    |
|----|-------|-------|-------|-------|-------|-------|-------|
| -3 | 0.67  |       | -0.18 | -0.43 |       | 0.02  |       |
| -2 |       | 0.56  |       |       | 0.31  |       | -0.40 |
| -1 | -0.18 |       | 0.17  | -0.08 |       | 0.31  |       |
| 0  | -0.43 |       | -0.08 | 0.31  |       | -0.29 |       |
| +1 |       | 0.31  |       |       | -0.12 |       | -0.08 |
| +2 | 0.02  |       | 0.31  | -0.29 |       | 0.54  |       |
| +3 |       | -0.40 |       |       | -0.08 |       | -0.07 |

**Trikontadipole-trikontadipole ($L=5$, $L'=5$)**

|    | -5    | -4    | -3    | -2    | -1    | 0     | +1    | +2    | +3    | +4    | +5    |
|----|-------|-------|-------|-------|-------|-------|-------|-------|-------|-------|-------|
| -5 | 0.36  |       | -0.16 |       | -0.07 | -0.48 |       | 0.52  |       | -0.13 |       |
| -4 |       | -0.06 |       | 0.01  |       |       | -0.08 |       | 0.11  |       | -0.13 |
| -3 | -0.16 |       | 0.20  |       | -0.03 | 0.06  |       |       |       | -0.20 |       |
| -2 |       | 0.01  |       | -0.04 |       |       | 0.05  |       | -0.07 |       | 0.08  |
| -1 | -0.07 |       | -0.03 |       |       |       |       | -0.03 |       | 0.09  |       |
| 0  | -0.48 |       | 0.06  |       |       | 0.05  |       | -0.14 |       |       |       |
| +1 |       | -0.08 |       | 0.05  |       |       | 0.22  |       | -0.10 |       | 0.63  |
| +2 | 0.52  |       |       |       | -0.03 | -0.14 |       | 0.30  |       | -0.17 |       |
| +3 |       | 0.11  |       | -0.07 |       |       | -0.10 |       | 0.01  |       | -0.28 |
| +4 | -0.13 |       | -0.20 |       | 0.09  |       |       | -0.17 |       | 0.21  |       |
| +5 |       | -0.13 |       | 0.08  |       |       | 0.63  |       | -0.28 |       | 0.35  |

**Dipole-octupole (*L*=1, *L'*=3)**

|     | -3    | -2   | -1    | 0    | +1    | +2    | +3   |
|-----|-------|------|-------|------|-------|-------|------|
| -1  | -0.34 |      | 0.08  | 0.40 |       | -0.04 |      |
| 0   | -0.34 |      | -0.21 | 0.22 |       | -0.27 |      |
| +1  |       | 0.08 |       |      | 0.05  |       | -0.07|

**Dipole-trikontadipole (*L*=1, *L'*=5)**

|     | -5    | -4    | -3    | -2    | -1    | 0     | +1    | +2    | +3    | +4    | +5   |
|-----|-------|-------|-------|-------|-------|-------|-------|-------|-------|-------|------|
| -1  | 0.07  |       | -0.30 |       | 0.02  | -0.09 |       | 0.03  |       | 0.19  |      |
| 0   | -0.05 |       | -0.19 |       | 0.09  | -0.10 |       | -0.13 |       | 0.27  |      |
| 1   |       | -0.07 |       | 0.04  |       |       | 0.11  |       | -0.05 |       | 0.13 |

**Octupole-trikontadipole (*L*=3, *L'*=5)**

|     | -5    | -4    | -3    | -2    | -1    | 0     | +1    | +2    | +3    | +4    | +5   |
|-----|-------|-------|-------|-------|-------|-------|-------|-------|-------|-------|------|
| -3  | -0.36 |       | 0.40  |       | -0.04 | 0.16  |       | -0.11 |       | -0.26 |      |
| -2  |       | -0.11 |       | 0.06  |       |       | 0.50  |       | -0.40 |       | 0.66 |
| -1  | 0.38  |       | -0.10 |       |       | -0.25 |       | 0.25  |       | -0.14 |      |
| 0   | 0.12  |       | -0.24 |       | 0.13  | -0.09 |       | -0.10 |       | 0.23  |      |
| +1  |       | -0.06 |       | 0.13  |       |       | 0.18  |       | -0.04 |       | 0.06 |
| +2  | 0.38  |       | 0.11  |       | -0.06 | -0.21 |       | 0.44  |       | -0.38 |      |
| +3  |       | 0.16  |       | -0.02 |       |       | -0.11 |       |       |       | -0.21|

## Charge SEIs

**Quadrupole-quadrupole (*L*=2, *L'*=2)**

|     | -2    | -1    | 0     | +1    | +2    |
|-----|-------|-------|-------|-------|-------|
| -2  | -0.14 |       |       | -0.13 |       |
| -1  |       | -0.19 | 0.01  |       | -0.02 |
| 0   |       | 0.01  | -0.42 |       | -0.49 |
| +1  | -0.13 |       |       | -0.14 |       |
| +2  |       | -0.02 | -0.49 |       | 0.15  |

**Hexadecapole-hexadecapole (*L*=4, *L'*=4)**

|     | -4    | -3    | -2    | -1    | +0    | +1    | +2    | +3    | +4    |
|-----|-------|-------|-------|-------|-------|-------|-------|-------|-------|
| -4  | -0.15 |       | 0.12  |       |       | -0.06 |       | 0.04  |       |
| -3  |       | 0.60  |       | -0.22 | -0.21 |       | 0.32  |       | -0.45 |
| -2  | 0.12  |       | -0.40 |       |       | 0.02  |       |       |       |
| -1  |       | -0.22 |       | 0.09  | 0.12  |       | -0.16 |       | 0.12  |
| 0   |       | -0.21 |       | 0.12  | 0.09  |       | -0.03 |       | 0.19  |
| 1   | -0.06 |       | 0.02  |       | -0.10 |       |       |       |       |
| 2   |       | 0.32  |       | -0.16 | -0.03 |       | -0.08 |       | -0.05 |
| 3   | 0.04  |       |       |       |       |       | -0.45 |       |       |
| 4   |       | -0.45 |       | 0.12  | 0.19  |       | -0.05 |       | 0.09  |

**Quadrupole-hexadecapole (*L*=2, *L'*=4)**

|     | -4    | -3    | -2    | -1    | +0    | +1    | +2    | +3    | +4    |
|-----|-------|-------|-------|-------|-------|-------|-------|-------|-------|
| -2  | 0.13  |       | -0.14 |       |       | -0.05 |       |       |       |
| -1  |       | -0.12 |       | 0.05  | 0.08  |       | -0.08 |       | 0.13  |
| 0   |       | -0.45 |       | 0.18  | 0.17  |       |       |       | 0.09  |
| 1   | 0.05  |       | 0.02  |       |       | -0.07 |       | 0.18  |       |
| 2   |       | 0.79  |       | -0.29 | -0.26 |       | -0.04 |       | -0.22 |

## 3. Contributions of dipolar and multipolar order parameters into the magnetic ordering energy

In order to quantify the role of various order parameters in the stabilization of magnetic order in $PrO_2$ we calculated the contributions of various interaction terms into its mean-field energy as

$$E_{LL'}^{MM'} = \frac{1}{N_M} \sum_i \sum_{j \in NN_i} V_{LM,L'M'}^{ij} \langle \hat{O}_{L(i)}^M \rangle \langle \hat{O}_{L'(j)}^{M'} \rangle,$$

where $N_M = 8$ is the number of Pr sites in the magnetic unit cell, $i$ runs over those sites, $j$ runs over the nearest neighbors of a given site $i$. The results are summarized in Table S2 below. One sees that the largest contributions stem from the SEIs between the $\hat{O}_3^1$ octupole and $\hat{O}_5^1$ trikontadipole.

| $V_{LM,L'M'}^{ij}$ | $\hat{O}_1^{-1}$ | $\hat{O}_1^1$ | $\hat{O}_3^{-1}$ | $\hat{O}_3^{-3}$ | $\hat{O}_3^1$ | $\hat{O}_3^3$ | $\hat{O}_5^{-3}$ | $\hat{O}_5^{-1}$ | $\hat{O}_5^1$ | $\hat{O}_5^3$ | $\hat{O}_5^5$ |
|---|---|---|---|---|---|---|---|---|---|---|---|
| $\hat{O}_1^{-1}$ | 0.03 | 0. | -0.69 | 0.43 | 0. | 0. | -0.07 | -0.14 | 0. | 0. | 0. |
| $\hat{O}_1^1$ | 0. | -0.04 | 0. | 0. | -0.83 | 0. | 0. | 0. | -0.34 | -0.13 | -0.08 |
| $\hat{O}_3^{-1}$ | -0.69 | 0. | -1.4 | 0.06 | 0. | 0. | -0.71 | 0.49 | 0. | 0. | 0. |
| $\hat{O}_3^{-3}$ | 0.43 | 0. | 0.06 | 0.23 | 0. | 0. | 0.19 | -0.72 | 0. | 0. | 0. |
| $\hat{O}_3^1$ | 0. | -0.83 | 0. | 0. | -4.85 | -0.06 | 0. | 0. | -2.84 | -0.28 | -0.24 |
| $\hat{O}_3^3$ | 0. | 0. | 0. | 0. | -0.06 | -0.04 | 0. | 0. | -0.28 | 0.05 | 0.01 |
| $\hat{O}_5^{-3}$ | -0.07 | 0. | -0.71 | 0.19 | 0. | 0. | -0.19 | -0.2 | 0. | 0. | 0. |
| $\hat{O}_5^{-1}$ | -0.14 | 0. | 0.49 | -0.72 | 0. | 0. | -0.2 | 0.57 | 0. | 0. | 0. |
| $\hat{O}_5^1$ | 0. | -0.34 | 0. | 0. | -2.84 | -0.28 | 0. | 0. | -1.58 | -0.01 | -0.19 |
| $\hat{O}_5^3$ | 0. | -0.13 | 0. | 0. | -0.28 | 0.05 | 0. | 0. | -0.01 | -0.05 | 0. |
| $\hat{O}_5^5$ | 0. | -0.08 | 0. | 0. | -0.24 | 0.01 | 0. | 0. | -0.19 | 0. | 0. |

Table S2: Contributions of the pair superexchange interactions between respective multipolar order parameters to the total energy of the ordered state of $PrO_2$ at zero temperature. The energies are in Kelvin per Pr site.

## 4. Calculated intensities of magnetic Bragg peaks

The calculate intensities of magnetic Bragg peaks are listed in Table S3 below. The calculations are carried out for the preferable structural domain, in which the field is directed along the *xy* [110] direction of the basal plane.

| *Reflection* | *(100)* | *(011)* | *(211)* | *(122)* | *(300)* |
|---|---|---|---|---|---|
| 0 T | 1.0 | 0.496 | 0.655 | 0.422 | 0.641 |
| 5 T | 0.996 | 0.000 | 0.512 | 0.075 | 0.636 |

Table S3: calculated intensities $I(\mathbf{q})$ of magnetic Bragg peaks in $PrO_2$ at $T$=1.5 K. The intensities are shown for the external magnetic field of 0 and 5 Tesla applied along the $[0\bar{1}1]$ cubic direction. All intensities are normalized to the intensity of (100) reflection in zero field.